\documentclass[aps,prl,twocolumn,superscriptaddress,showpacs,preprintnumbers,amsmath,amssymb]{revtex4}

\usepackage{orcidlink} 

\usepackage{afterpage,rotating}
\usepackage{currvita}

\usepackage{tabularx}
\usepackage{nonfloat}
\usepackage{graphicx}
\usepackage{dcolumn}
\usepackage{bm}
\usepackage{subfigure}
\usepackage{mathrsfs}
\usepackage{amsmath,amsbsy,amssymb,lmodern}
\usepackage{epsfig,color}
\usepackage{placeins}
\usepackage{rccol}
\usepackage{comment}
\usepackage{hyperref} 
\usepackage{setspace} 
\usepackage{tabularx} 
\usepackage{multirow}
\usepackage{braket}
\usepackage{accents}

\graphicspath{{ps}}

\usepackage{xcolor}
\usepackage{enumitem}

\newcommand{\bseep}{B_s^0 \rightarrow \eta^\prime \eta}

\DeclareRobustCommand\mybar[1]{\accentset{\rule{0.5em}{0.6pt}}{#1}}

\renewcommand{\arraystretch}{1.1}

\usepackage{helvet}

\begin{document}
%

%



\preprint{\vbox{ \hbox{ }
\hbox{Belle Preprint 2024-04 \hfill}
\hbox{KEK Preprint 2024-5 \hfill}
}}


\title{ \quad\\[1.0cm]
Search for Two-Body $B$ Meson Decays to $\Lambda^{0}$ and $\Omega^{(*)0}_{c}$
}





\noaffiliation
  \author{V.~Savinov\,\orcidlink{0000-0002-9184-2830}} 
  \author{I.~Adachi\,\orcidlink{0000-0003-2287-0173}} 
  \author{J.~K.~Ahn\,\orcidlink{0000-0002-5795-2243}} 
  \author{H.~Aihara\,\orcidlink{0000-0002-1907-5964}} 
  \author{D.~M.~Asner\,\orcidlink{0000-0002-1586-5790}} 
  \author{H.~Atmacan\,\orcidlink{0000-0003-2435-501X}} 
  \author{R.~Ayad\,\orcidlink{0000-0003-3466-9290}} 
  \author{Sw.~Banerjee\,\orcidlink{0000-0001-8852-2409}} 
  \author{J.~Bennett\,\orcidlink{0000-0002-5440-2668}} 
  \author{M.~Bessner\,\orcidlink{0000-0003-1776-0439}} 
% ...
\collaboration{The Belle Collaboration}

\begin{abstract}
\par We report the results of the first search for 
Standard Model and baryon-number-violating 
two-body decays
of the neutral $B$ mesons to
$\Lambda^{0}$
and
$\Omega^{(*)0}_c$
using 711~${\rm fb^{-1}}$ of data
collected at the $\Upsilon(4S)$ resonance
with the Belle detector
at the KEKB asymmetric-energy $e^+ e^-$ collider.
We observe no evidence of signal from any such decays
and set 95\% confidence-level upper limits
on the products of
$B^0$ and $\mybar{B}^0$ branching fractions
for these two-body decays
with $\mathcal{B}(\Omega_{c}^{0} \to \pi^+ \Omega^-)$
in the range between 9.5~$\times 10^{-8}$ and 31.2~$\times 10^{-8}$.

\end{abstract}




\pacs{11.30.Fs, 13.30.-a, 13.20.He}

\maketitle

\tighten

{\renewcommand{\thefootnote}{\fnsymbol{footnote}}}
\setcounter{footnote}{0}


%

\begin{figure}[!ht]
\begin{center}
\begin{minipage}[c]{0.95\linewidth}
\small
 \begin{center}
    \subfigure{\includegraphics[width=1.0\textwidth]{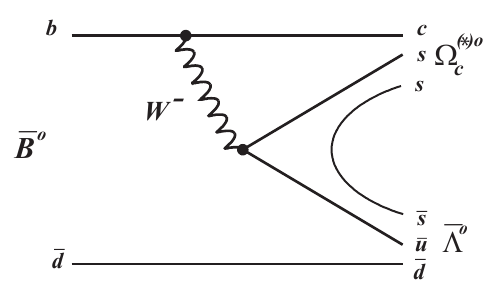}}
 \end{center}
\end{minipage}\hfill
\\
\begin{minipage}[c]{0.95\linewidth}
\small
 \begin{center}
    \subfigure{\includegraphics[width=1.0\textwidth]{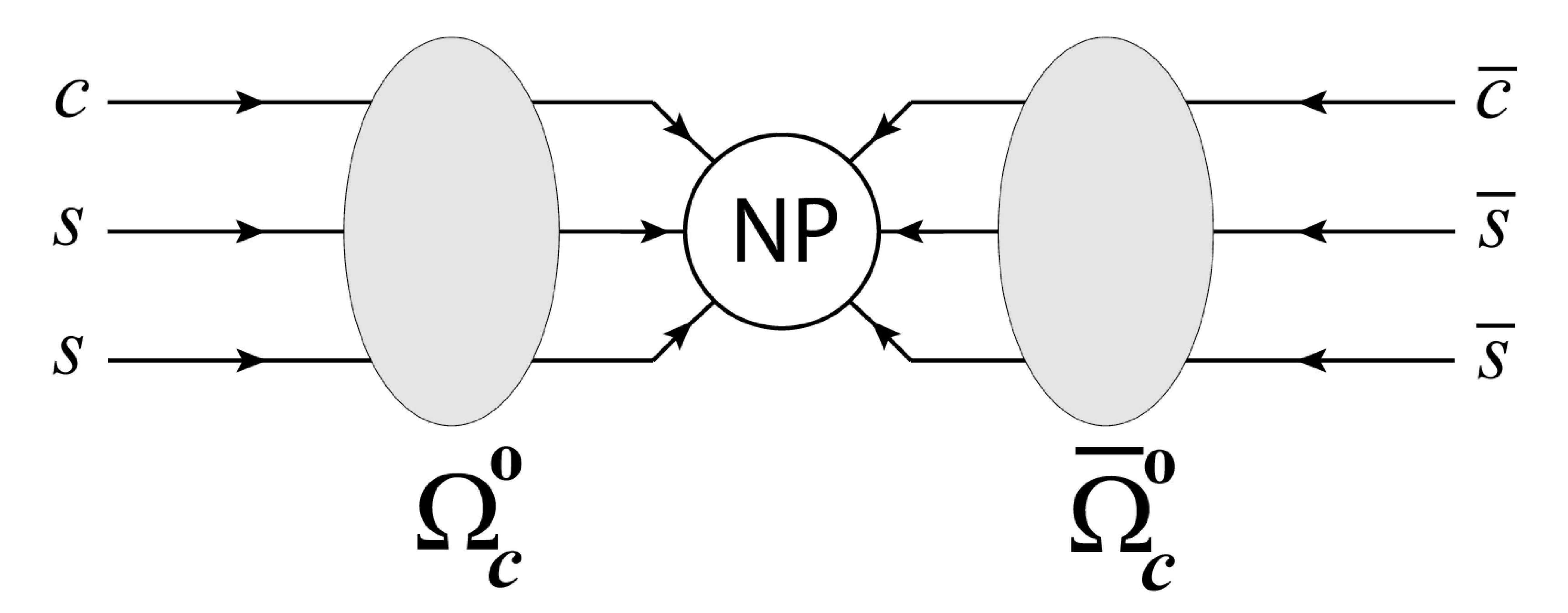}}
 \end{center}
\end{minipage}\hfill
\end{center}
\figcaption{
Quark-level Feynman diagram for signal SM decays
and
a depiction of the
$\Omega_{c}^{0} - \mybar{\Omega}_{c}^{0}$ oscillation hypothesis.
}
\label{fig_1}
\end{figure}

In the analysis presented in this Letter 
we search for new two-body $B$ decays
to final states with
$\Lambda^{0}$
and
$\Omega^{0}_c$,
where Beyond the Standard Model~(BSM) amplitudes could contribute directly 
or
as the result of a Standard Model~(SM) decay followed by baryon-antibaryon oscillations
of $\Omega_{c}^{0}$ or $\Lambda^{0}$. 
Our analysis includes the SM Cabibbo-suppressed decay 
$\mybar{B}^{0} \to \mybar{\Lambda}^{0} \Omega_{c}^{(*)0}$ 
and 
the BSM decay 
$\mybar{B}^{0} \to \mybar{\Lambda}^{0} \mybar{\Omega}_{c}^{(*)0}$. 
The former SM decay proceeds via the $b \to c$ tree transition 
and is poorly understood from 
a theoretical perspective 
because of hadronic uncertainties. 
The latter BSM  decay could result from 
the $\Omega_{c}^{0} - \mybar{\Omega}_{c}^{0}$ oscillations, 
a scenario 
which was suggested recently~\cite{Aitken:2017wie} 
as a low-energy mechanism for baryon number violation~(BNV),
which is one of the three Sakharov's 
conditions for baryogenesis~\cite{Sakharov:1967dj}. 
Additional BSM decays explored in our analysis include 
$\mybar{B}^{0} \to \Lambda^{0} \Omega_{c}^{(*)0}$, 
which is exceedingly unlikely in part due to 
the stringent limit recently set on 
$\Lambda^{0} - \mybar{\Lambda}^{0}$ oscillations 
by the BES~III experiment~\cite{BESIII:2023tge}. 
%
%
%
A Feynman diagram for a quark-level SM transition
investigated in our analysis
and
a depiction of the $\Omega^{0}_c - \mybar{\Omega}_{c}^{0}$ BSM oscillation hypothesis
are shown in Fig.~\ref{fig_1}.


%

To carry out the analysis described in this Letter 
we use the full Belle data sample of $711~\textrm{fb}^{-1}$
collected at the $\Upsilon(4S)$ resonance. 
This data sample contains $772$ million $B\bar{B}$ pairs~\cite{Belle:2012iwr}.
We search
for two-body decays
of the neutral $B$ mesons to
$\Lambda^{0}$
and
one of the two $\Omega^{0}_c$ states, 
$\Omega^{0}_c$ or $\Omega_{c}^{}(2770)^{0}$, 
previously known as $\Omega_{c}^{*0}$ 
(or their antiparticles). 
When we refer to either of these two $\Omega^{0}_c$ states, 
we use the notation $\Omega_{c}^{(*)0}$.
We reconstruct $\Omega_c^0$ in its decay to $\pi^+$ and $\Omega^-$, 
where $\Omega^-$ is detected in the $\Lambda^0 K^-$ channel.
The $\Lambda^0$ is reconstructed via $\pi^-$ and proton.
The decay of $B$ meson to
$\Lambda^{0}$ and $\Omega_c^{}(2770)^{0}$
is partially reconstructed:
the radiative photon from the decay
${\Omega_{c}^{}(2770)}^{0} \to \gamma \Omega_c^{0}$,
which is assumed to be the only decay of $\Omega_c^{}(2770)^{0}$,
is 
not explicitly reconstructed 
in the analysis.

Two final states,
$\mybar{\Lambda}^{0} \Omega^{0}_c$ and
$\mybar{\Lambda}^{0} \mybar{\Omega}^{0}_c$,
are studied in our analysis~\cite{cc} 
of the decays of the neutral $B$ mesons.
The observation of the signal
in the former final state
would indicate the SM $\mybar{B}^0$ decay,
while the latter final state
could be either
due to the former SM decay
followed by the $\Omega_{c}^{0} - \mybar{\Omega}_{c}^{0}$ oscillations
or due to
the direct BNV $\mybar{B}^0$ decay.
In this Letter,
the final states
$\mybar{\Lambda}^{0} \Omega^{0}_c$
and
$\mybar{\Lambda}^{0} \mybar{\Omega}^{0}_c$
are referred to as
SM-compatible
and
exclusively-BSM
channels,
respectively.
As we perform no $B$ flavor tagging,
any of the final states in this analysis
could be attributed to
$\mybar{B}^0$  or $B^0$; 
therefore, 
when discussing the signal processes,
we use the notation $B$
to refer to either
$B^0$ or $\mybar{B}^0$ mesons.
We
report the results
for the products of
$\mathcal{B}(B \to \mybar{\Lambda}^{0} \Omega^{0}_c)$,
$\mathcal{B}(B \to \mybar{\Lambda}^{0} \mybar{\Omega}^{0}_c)$,
$\mathcal{B}(B \to \mybar{\Lambda}^{0} \Omega_c^{}(2770)^{0})$ and
$\mathcal{B}(B \to \mybar{\Lambda}^{0} \mybar{\Omega}_c^{}(2770)^{0})$
with $\mathcal{B}(\Omega_{c}^{0} \to \pi^+ \Omega^-)$,
where the latter branching fraction,
quite interesting in its own right,
has not been measured yet.

%
%

The Belle detector~\cite{Belle} is
a large-solid-angle magnetic spectrometer
that operated at the KEKB asymmetric-energy $e^+e^-$ collider~\cite{KEKB}.
The detector components relevant to our study include
a silicon vertex detector~(SVD), a central drift chamber~(CDC),
a particle identification~(PID) system
that consists of a barrel-like arrangement of time-of-flight scintillation counters~(TOF)
and an array of aerogel threshold Cherenkov counters~(ACC),
and a CsI(Tl) crystal-based electromagnetic calorimeter~(ECL).
All these components are located inside a superconducting solenoid coil that
provides
a 1.5~T magnetic field.
%

To maximize discovery potential of the analysis
and
to validate the signal extraction procedure
we use a sample of simulated background events
referred to as generic Monte Carlo~(MC),
equivalent to six times the integrated luminosity
of the full Belle data sample.
These events simulate hadronic continuum,
i.e., quark-pair production in $e^+ e^-$ annihilation,
the decay of the $\Upsilon(4S)$ resonance into pairs of $B$ mesons
and the subsequent decays of the latter 
according to known branching fractions~\cite{PDG}.
Hadronic continuum represents
the main source of background in our analysis.
To estimate the overall reconstruction efficiency,
we use several
high-statistics signal MC samples,
where the non-signal $B$ meson decays generically.
We use the MC generator {\sc EvtGen}~\cite{EvtGen}
to simulate the production and decay processes
at and near the production point of
$\Upsilon(4S)$,
and
the {\sc GEANT} toolkit~\cite{GEANT}
to model detector response
and to handle the decays
of $\Omega^{-}$ and $\Lambda^{0}$.
To model final state radiation,
MC generator {\sc PHOTOS}~\cite{Barberio:1990ms}
is employed.
Hadronization is modeled using MC generator {\sc PYTHIA}~\cite{Sjostrand:2007gs}.
In this analysis 
we exclusively reconstruct
$\mybar{\Lambda}^{0} \Omega^{0}_c$ 
($\mybar{\Lambda}^{0} \mybar{\Omega}^{0}_c$)
final state using six charged particles:
three pions,
a kaon, 
and 
a proton and an antiproton 
(two antiprotons). 
The selection criteria are optimized
to suppress background in the SM-compatible channel
and to reduce
systematic uncertainties.
Reconstructed charged particles
are required to have $p_\perp$, 
the magnitude of the transverse part of their momenta with respect 
to the $z$ axis which is opposite to the direction of the $e^+$ beam, 
larger than 50~MeV/{$c$}. 
This requirement
removes candidates found in the region where
the efficiency has a large uncertainty and is very small.
The efficiency of this selection
for signal MC events is 28\%,
which is due to the kinematics of the signal process.
To select signal particle candidates of the correct species,
we apply requirements to the likelihood ratios,
$R_{s/r} = L_{s} /~(L_{s} + L_{r})$,
which are based
on PID measurements~\cite{Bevan:2014iga},
where $L_{s}$ and $L_{r}$ are the likelihoods
according to the $s$ and $r$ particle species hypotheses,
respectively.
The likelihood for each particle species is obtained
by combining information from CDC, TOF, ACC, and,
for the electron/hadron likelihood ratio $R_{e/h}$,
also ECL.
Our requirements are
$R_{\pi/K} \ge 0.6$ and $R_{e/h} \le 0.95$ for pion from $\Omega^{0}_c$ decay,
$R_{K/\pi} \ge 0.4$ and $R_{e/h} \le 0.95$ for kaon  from $\Omega^{-}$ decay,
and
$R_{p/K} \ge 0.1$ for protons.
The efficiency of PID requirements depends on the particle species
and kinematics, and varies between 92\% and 98\%.
PID misidentification rate is between 4\% and 6\%
per particle.
PID-based selection applied to all six
charged particles used to reconstruct
analyzed final states
rejects 85\% of background
and is 71\%-efficient for signal MC events.

Thus selected final state particles
are used to reconstruct
the following decays of signal baryon candidates:
$\Lambda^{0} \to p \pi^-$,
$\Omega^{-} \to K^- \Lambda^{0}$,
and
$\Omega^{0}_c \to \pi^+ \Omega^-$.
To identify the $\Lambda^{0}$ candidates,
we search for secondary vertices
associated with pairs of oppositely charged particles. 
To improve mass resolution and to suppress the combinatorial background,
the tracks reconstructed for these particles are refit to a common vertex.
The four-momenta obtained from this kinematic fit are used for further analysis.
The reconstructed mass of the $\Lambda^{0}$ candidate is required to be
within 8~MeV/$c^2$
(5.4~$\sigma$ of Gaussian resolution)
of the $\Lambda^{0}$ nominal mass~\cite{PDG}.
To reconstruct an $\Omega^{-}$ candidate,
we add together the four-momenta of $\Lambda^{0}$ and $K^-$ candidates
after refitting the $\Lambda^{0}$ vertex
while constraining its reconstructed mass
to the $\Lambda^{0}$ nominal mass.
The reconstructed mass of the $\Omega^{-}$ candidate is required to be
within 60~MeV/$c^2$~(15.1~$\sigma$) of the $\Omega^-$ nominal mass~\cite{PDG}.
Then the $\Omega^-$ candidate
undergoes a procedure
similar to the one used for $\Lambda^{0}$ candidates:
we perform a vertex fit, update kinematics of the daughter particles, and
require the reconstructed mass of the $\Omega^{-}$ candidate
to be within 7~MeV/$c^2$~(4.4~$\sigma$) of the $\Omega^{-}$ nominal mass~\cite{PDG},
then we repeat a vertex fit with mass constraint.
To suppress the combinatorial background,
the distance between the interaction point and 
the decay vertex of $\Omega^{-}$,  
i.e., decay length is required to be greater than 0.5~cm~(6.5~$\sigma$).
This selection reduces background by 43\% and,
using SM signal channel with $\Omega^{0}_c$ as a reference,
keeps 83\% of the signal.
The procedure for reconstructing $\Omega_{c}^{0}$ candidates is similar
to the one used to reconstruct $\Omega^{-}$
when one replaces $\Lambda^{0}$ by $\Omega^-$ and $K^-$ by $\pi^+$,
as well as requiring the invariant mass to be within 100~MeV/$c^2$~(17.5~$\sigma$)
before the vertex fit and 19~MeV/$c^2$~(4.1~$\sigma$) after the vertex fit.
The per-degree-of-freedom $\chi^2$ from each kinematic fit
in the described procedure
is required to be less than 100.
The overall efficiency of the vertex-mass $\chi^2$-based requirements is 83\%,
while the background is suppressed
by a factor of approximately 10.
The requirements applied to the reconstructed invariant masses
after kinematic fitting are 91\%-efficient
and remove 89\% of the remaining background.
 
\begin{figure}[!ht]
\small
\begin{minipage}[c]{0.50\linewidth}
\small
 \begin{center}
    \subfigure{\includegraphics[width=1.0\textwidth]{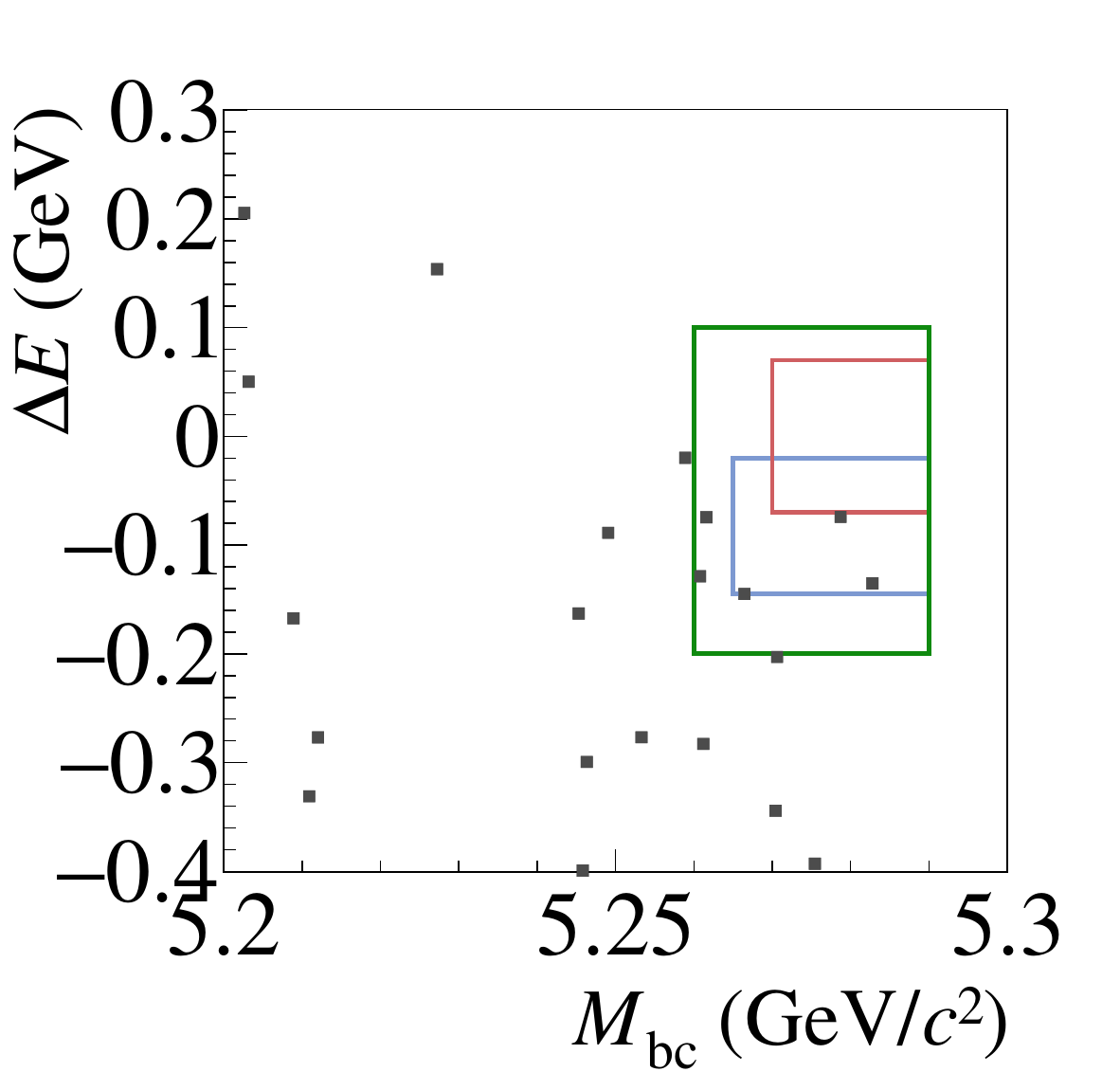}}
 \end{center}
\end{minipage}\hfill
\begin{minipage}[c]{0.50\linewidth}
\small
 \begin{center}
    \subfigure{\includegraphics[width=1.0\textwidth]{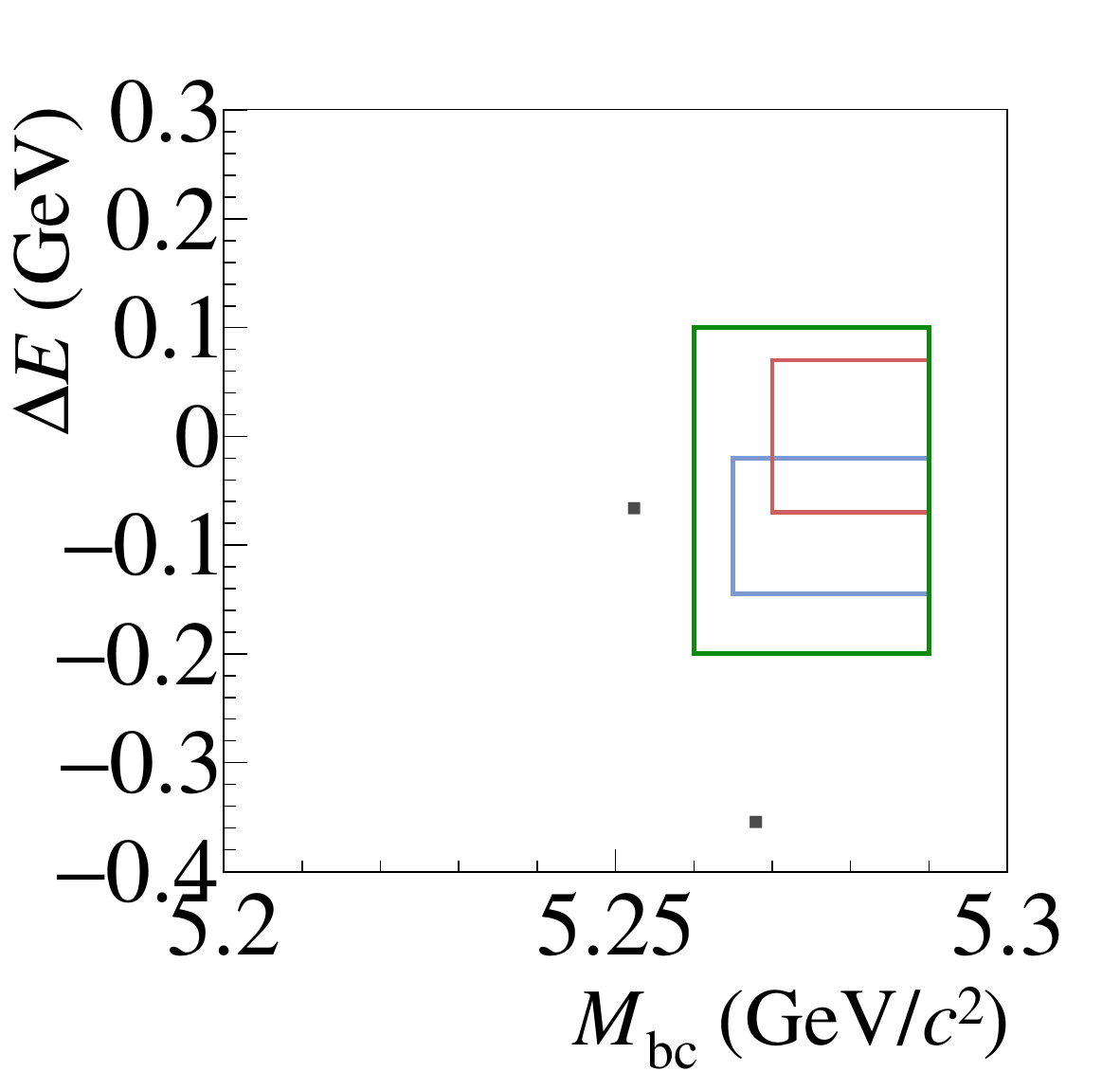}}
 \end{center}
\end{minipage}\hfill
\figcaption{
The distributions of the best candidates
in data events
after applying all selection criteria
in the analyzed region of
$\Delta E$
vs
$M_{\rm bc}$
for
SM-compatible
$\mybar{\Lambda}^{0} \Omega^{0}_c$~(left)
and
exclusively-BSM
$\mybar{\Lambda}^{0} \mybar{\Omega}^{0}_c$~(right)
reconstructed final states.
The candidates are shown with square markers.
The smaller boxes
with red and blue outlines
show the signal regions for final states with
$\Omega^{0}_c$
and
$\Omega_{c}^{}(2770)^{0}$,
respectively.
The larger box with a green outline
represents the separation between
the blinded region
and
the sideband.
}
\label{fig_2}
\end{figure}

To reconstruct signal $B$ meson candidates, 
we use beam-energy-constrained $B$ mass 
$M_{\rm bc} \equiv \sqrt{E_{\rm beam}^2/c^4 - p_{B}^2/c^2}$
and the energy difference
$\Delta E \equiv E_{B} - E_{\rm beam}$,
where $p_{B}$, $E_{B}$, and $E_{\rm beam}$ are
the momentum and energy of the $B$ candidate,
and the beam energy,
respectively,
evaluated in the $e^+ e^- $ center-of-mass frame. 
The momentum and energy of the $B$ candidate 
are evaluated by adding together the four-momenta of
$\Lambda^{0}$ and $\Omega^{0}_c$ candidates 
after vertex fits with mass constraints.
No vertex fit is performed for the $B$ candidate,
as doing so is found to have
no tangible benefit for background suppression
and separation between the $\Omega_c^{0}$
and $\Omega_c^{}(2770)^{0}$ signals.
%
%
We require
$M_{\rm bc} \, > \, 5.200 {\rm ~GeV}/c^2$
and
$-400 \, {\rm MeV} \, \le \, \Delta E \, \le \, 300 \, {\rm MeV}$.
The efficiency of these last two selection criteria exceeds 99\%.
%
%
We define signal regions
for final states with $\Omega^{0}_c$ to be
$M_{\rm bc} \, > \, 5.270 {\rm ~GeV}/c^2$
and
$-70 \, {\rm MeV} \, \le \, \Delta E \, \le \, 70 \, {\rm MeV}$,
and
for final states with $\Omega_c^{}(2770)^{0}$
(where we ignore the radiative photon from the decay $\Omega_c^{}(2770)^{0} \to \gamma \Omega^{0}_c$)
to be
$M_{\rm bc} \, > \, 5.265 {\rm ~GeV}/c^2$
and
$-145 \, {\rm MeV} \, \le \, \Delta E \, \le \, -20 \, {\rm MeV}$.
%
%
Each of the two signal regions contains at least 98\% of the respective signal.
A slightly larger region,
$M_{\rm bc} \, > \, 5.260 {\rm ~GeV}/c^2$
and
$-200 \, {\rm MeV} \, \le \, \Delta E \, \le \, 100 \, {\rm MeV}$,
which includes the union of the two signal regions,
is blinded.
The non-blinded region of $M_{\rm bc}$ and $\Delta E$ defines the sideband.
The defined regions are illustrated in Fig.~\ref{fig_2}.

When events contain more than one candidate
(which occurs $4.2\%$ of the time in signal MC,
corresponding to the average candidate multiplicity of 1.05),
we select the best candidate according
to the smallest value of cumulative $\chi^2$
obtained from the four vertex fits with mass constraints.
Multiple candidates in this analysis
are usually associated with particles
of relatively low $p_\perp$,
when multiple tracks with similar parameters
are reconstructed for the same charged particle.
We study signal MC events to prove that
our procedure selects candidates
with the best invariant mass
and $\Delta E$ resolutions.
Simulation demonstrates that,
for events with multiple candidates,
the best candidate selected by our method
has all 6 signal particles reconstructed correctly
in 93\% of signal MC events.
The best candidate is selected
after applying all other analysis requirements.
While the data and generic MC clearly contain $\Lambda^{0}$ and $\Omega_c^0$ mesons,
the distributions of $M_{\rm bc}$ and $\Delta E$
for simulation exhibit only non-peaking background,
which is of combinatorial origin.

The background suppression 
and analysis sensitivity 
optimization 
are performed
using the signal region
for SM-compatible channel with $\Omega_c^{0}$.
%
To suppress combinatorial background, 
the $\Omega^-$ decay length requirement
is optimized by maximizing $S/\sqrt{B}$ figure of merit~(FOM), 
where $S$ and $B$ are the numbers of signal and background
MC events satisfying this requirement. 
To provide better sensitivity 
to the SM signal, 
the requirements applied 
to the reconstructed invariant masses 
of $\Lambda^{0}$, $\Omega^-$ and $\Omega_c^0$ 
after the vertex fits
are optimized by maximizing
the value of Punzi's FOM~\cite{punzi}:
%
$\varepsilon(t)/(a/2+\sqrt{B(t)})$,
%
where
$\varepsilon(t)$ and $B(t)$ are the signal reconstruction efficiency
and the number of background events expected in the signal region
for a given set of requirements, $t$
(applied to the reconstructed invariant masses),
respectively. 
The quantity $a$ is the desired significance 
in units of standard deviation. 
To predict $B(t)$, we multiply
the number of events in the data sideband by
the ratio of the numbers of events
in the signal region and sideband
in our generic MC sample.
We require the optimized values 
of these selection criteria 
to be at least 4~units of Gaussian resolution 
away from the nominal masses~\cite{PDG}. 
This leads us to use $a=10$. 
%
%
The choices made for the rest of the selection criteria 
represent a balance between maximizing the efficiency 
and minimizing the systematic uncertainties. 
%
%
%
The overall detection efficiencies for individual channels 
are in the range between 11.5\% and 12.4\%. 

After applying all selection criteria, 
the total numbers of events remaining in data
outside the blinded regions
for final states
$\mybar{\Lambda}^0 \Omega^{0}_c$
and
$\mybar{\Lambda}^0 \mybar{\Omega}^{0}_c$
are
16 and 2 events, respectively.
Using MC simulation,
we estimate that both
hadronic continuum
and
non-signal $B\bar{B}$ events
contribute to the sideband.
Using sideband data and
the scaling factor of
$0.10 \pm 0.04$
($0.09 \pm 0.06$)
obtained from generic MC,
we expect to find
$1.6 \pm 0.7$
($0.18 \pm 0.17$)
background events in the blinded region
for SM-compatible~(exclusively-BSM) channels. 
%
%

%
%

Upon unblinding, 
in addition to the 16 events in the sideband, 
we find 5 events 
in the 
blinded region 
for the SM-compatible channels 
(of which 0 events are 
in the $\mybar{\Lambda}^{0} \Omega_{c}^{0}$ signal region
and 3 events are
in the $\mybar{\Lambda}^{0} \Omega_{c}^{}(2770)^{0}$ signal region). 
In the exclusively-BSM channels,
2 events are observed,
both outside the blinded region.
The numbers of events observed in data are shown in Table~\ref{table_events}.
The 2D distributions of $\Delta E$ vs $M_{\rm bc}$
for the best candidates in data events are shown in Fig.~\ref{fig_2}.
The observation of 5 events in the SM blinded region including 3 events
in the $\mybar{\Lambda}^{0} \Omega_{c}^{}(2770)^{0}$ signal region 
is consistent with SM background expectations. 
Statistical significance of such background fluctuation is less than 3 standard deviations. 
When we estimate the upper limits 
we assume that no background is expected in the signal region. 

{\renewcommand{\arraystretch}{1.2}
\begin{center}
\begin{table}[htb!]
\begin{tabular}{c|c|c|c|c}
\hline
Numbers             &       &          & $\mybar{\Lambda}^{0} \Omega_{c}^{0}$ & $\mybar{\Lambda}^{0} \Omega_{c}^{}(2770)^{0}$ \\
of                  & Total & Blinded  & signal                               & signal                                     \\
events              &       & region   & region                               & region                                     \\
\hline
$\mybar{\Lambda}^{0} \Omega_{c}^{0}$ data         & 21 &  5 & 0 & 3 \\
Background                                & N/A & $1.6 \pm 0.7$ & $0.44 \pm 0.45$ & $0.44 \pm 0.45$  \\
\hline
$\mybar{\Lambda}^{0} \mybar{\Omega}_{c}^{0}$ data &  2 &  0 & 0 & 0 \\
Background                                & N/A & $0.18 \pm 0.17$ & $0.00 \pm 0.12$ & $0.12 \pm 0.15$ \\
\hline
\end{tabular}
\hfill
\tabcaption{
Numbers of events observed in data and background estimates.
}
\label{table_events}
\end{table}
\end{center}
}

Systematic uncertainties arise from imprecise knowledge
of various efficiencies and other quantities detailed in Table~\ref{table_syst}.
We assign a 0.35\% uncertainty for tracks reconstructed for charged particles
with $p_\perp >$~200~MeV/{\it c}~\cite{Belle:2014mfl}
and
a 1.2\% uncertainty for particles
with 50~MeV/{\it c} $\leq p_\perp \leq$~200~MeV/{\it c}~\cite{Belle:2011cxw}.
Therefore, based on the fractions of signal MC particles in these two momentum regions,
we assign 2.9\% to total uncertainty for all six tracks.
Uncertainties due to PID selection for charged kaons and pions
are obtained from high-statistics comparisons between MC and data~\cite{Nakano:2002jw}.
As no dedicated study has been performed
for the 98\%-efficient proton PID requirement $R_{p/K} \geq 0.1$,
we assign a 2\% uncertainty on basis of its high efficiency.
To estimate the magnitudes of possible
MC-data differences in 
the efficiencies 
of selection criteria applied to
the $\Omega^-$ decay length,
$\chi^2$ from kinematic fits
and
the reconstructed masses,
we vary selection criteria within ranges
typical for Belle analyses as described below.
Uncertainties for the requirements
applied to the $\Omega^-$ decay length
and reconstructed masses are determined
by varying the values of selection criteria by $\pm 10$\%
and using the change in the efficiency
as an estimate of systematic uncertainty.
Similarly, the systematic uncertainty in the efficiency
of the vertex-mass kinematic fits $\chi^2$ criteria
is estimated as the larger change in reconstruction efficiency for signal MC
when the $\chi^2$ requirement is varied by  $\pm 25$ with respect to the nominal value of 100.
Total uncertainty for $M_{\rm bc}$ and $\Delta E$ selection criteria is estimated to be 0.5\%
which is half of their inefficiency.
Uncertainties 
on daughter branching fractions, 
$\mathcal{B}(\Omega^{-} \! \to \Lambda^{0} K^{-}) = 0.678$ 
and 
$\mathcal{B}(\Lambda^{0} \to p \pi^{-}) = 0.641$,
and 
on the number of $B^0 \mybar{B}^0$ pairs in Belle data, 
$N_{B^0\mybar{B}^0} = 374 \times 10^{6}$, 
are taken from References \cite{PDG} and ~\cite{Belle:2012iwr,Belle:2022hka},
respectively.
Relative difference between efficiencies 
for the charge conjugate final states in MC 
is approximately 0.4\%. 
This detector charge asymmetry arises 
because of the difference in the detector response to particles and antiparticles. 
While simulation is known to underestimate this effect, 
in our analysis its impact is reduced because 
we combine charge conjugate final states in efficiency estimates. 
To account for limitations of our approach, 
a relative systematic uncertainty of 0.8\% is assigned to the efficiencies, 
which is twice of the MC-predicted difference between efficiencies 
of individual charge conjugate states. 
Possible presence of BSM amplitudes could affect angular distributions of baryons in the signal decay chain.
This effect is likely to be small in decays of pseudoscalar $B$ meson. To be conservative,
we assign relative systematic uncertainty of 0.5\% based on MC studies of efficiency's dependence
on helicity angles, i.e., baryons' polarization.
Finally, relative uncertainty due to MC statistics is determined using the formula
$\sqrt{\epsilon \times~(1-\epsilon)/N} \times \left(1/\epsilon\right)$,
where $\epsilon$ is the overall signal reconstruction efficiency and $N$ is
the number of signal MC events~(before reconstruction).
We combine 
the uncertainties for all contributions 
in quadrature, 
while taking into account also 
the 100\% correlation 
between two $\Lambda^0 \to p \pi^-$ decays,
two protons 
and 
correlated selection criteria for reconstructed masses, 
to estimate overall systematic uncertainty $\sigma_r$ to be 6.2\%.

{\renewcommand{\arraystretch}{1.2}
\begin{center}
\begin{table}[htb!]
\begin{tabular}{l|c}
\hline
Source & Uncertainty~(\%)  \\
\hline
Track reconstruction~(overall) & 2.9 \\
$\pi^+$ PID~(for $\Omega^{0}_c \to \pi^+ \Omega^-$) & 0.8 \\
$K^-$ PID~(for $\Omega^- \to K^- \Lambda^0$) & 1.4 \\
$p$ PID~(for $\Lambda^0$ decays) & $2 \times 1.0$ \\
Decay length~($\Omega^-$) & 2.0 \\
Reconstructed masses & $4 \times 0.5$ \\
Vertex fits~($\chi^2$) & 1.5 \\
$M_{\rm bc}$ and $\Delta E$ & 0.5 \\
$\mathcal{B}(\Omega^- \to \Lambda^0 K^-)$ & 1.0 \\
$\mathcal{B}(\Lambda^0 \to p \pi^-)$ & $2 \times 0.7$ \\
$N_{B^0\bar{B}^0}$ & 2.9 \\
Detector charge asymmetry & 0.8 \\
Polarization of baryons & 0.5 \\
MC statistics & 0.7 \\
\hline
Overall~($\sigma_r$) & 6.2 \\
\hline
\end{tabular}
\hfill
\tabcaption{{Summary of relative systematic uncertainties.}}
\label{table_syst}
\end{table}
\end{center}
}

%
%
%
%
%
We estimate the 95\%~CL upper limits 
on the products of branching fractions 
$
\mathcal{B}(B \to \mybar{\Lambda}^{0} \Omega_{c}^{(*)0})
\times
\mathcal{B}(\Omega_{c}^{0} \to \Omega^{-} \pi^{+}) 
$
as 
$
U_{n}/
(2
\times
N_{B^0\bar{B}^0}
\times
\epsilon
\times
\mathcal{B}(\Omega^{-} \to \Lambda^{0} K^{-})
\times
\mathcal{B}(\Lambda^{0} \to p \pi^{-})^{2}
),
$
%
\noindent where
$B$ is assumed to be solely either $B^0$ or $\mybar{B}^0$, 
and 
%
%
$U_n$ is the 95\% CL upper limit 
on the number of events. 
Systematic uncertainties 
are included 
using the approach by Cousins and Highland~\cite{CousinsHighland:1992ch}: 
%
$U_n = U_{n0}\left(1 +~(U_{n0} - n) \sigma_r^2/2\right)$, 
%
\noindent 
where 
$n$ is the number of events observed in data in the signal region~(including charge-conjugate final state). 
To estimate $U_{n0}$, the 95\% CL upper limit on the number of events in data without systematic uncertainties, 
we use the upper bounds of the 90\% CL intervals tabulated in the seminal paper by Feldman and Cousins~\cite{fc}.
When using these confidence intervals, 
we assume that no background events are expected in either signal region. 
For the channel where 3 events are observed,
we extend the lower bound of the respective CL interval to 0. 
This procedure is motivated by 
a large uncertainty in background estimates 
and 
results in overcoverage. 
%
%
Various quantities necessary to complete the calculations are shown
in Tables~\ref{table_syst} and~\ref{table_calc}.
Following the procedure
described above,
we summarize
the resulting 95\% CL upper limits
in Table~\ref{table_results}.
The upper limits on the products of branching fractions 
$\mathcal{B}(B \to \mybar{\Lambda}^{0} \mybar{\Omega}_{c}^{(*)0})
\times
\mathcal{B}(\Omega_{c}^{0} \to \Omega^{-} \pi^{+})$
are estimated similarly. 
%

%
%
%
%
%
%

{\renewcommand{\arraystretch}{1.2}
\begin{center}
\begin{table}[htb!]
\begin{tabular}{l|cccc}
\hline
Channel &
$\mybar{\Lambda}^{0} \Omega_{c}^{0}$
&
$\mybar{\Lambda}^{0} \Omega_{c}^{}(2770)^{0}$
&
$\mybar{\Lambda}^{0} \mybar{\Omega}_{c}^{0}$
&
$\mybar{\Lambda}^{0} \mybar{\Omega}_{c}^{}(2770)^{0}$
\\
\hline
$\epsilon$~(\%) & 12.1 & 11.5 & 12.4 & 11.8 \\
$n$~(events) & 0 & 3 & 0 & 0 \\
$U_{n0}$~(events)& 2.44 & 7.42 & 2.44 & 2.44 \\
$U_{n}$~(events)& 2.45 & 7.48 & 2.45 & 2.45 \\
\hline
\end{tabular}
\hfill
\tabcaption{Information needed for upper limit estimates.}
\label{table_calc}
\end{table}
\end{center}
}

{\renewcommand{\arraystretch}{1.2}
{
\begin{table}
\begin{ruledtabular}
\begin{tabular}{c|c}
Quantity~($\times \mathcal{B}(\Omega_{c}^{0} \to \Omega^{-} \pi^{+})$)     & Upper limit~(at 95\% CL) \\
\hline
\small{\multirow{2}{*}{$\mathcal{B}(B \to \mybar{\Lambda}^{0} \Omega_{c}^{0})$}} &  \small{\multirow{2}{*}{$9.7 \times 10^{-8}$}} \\
 & \\

\hline
\small{\multirow{2}{*}{$\mathcal{B}(B \to \mybar{\Lambda}^{0} \Omega_{c}^{}(2770)^{0})$}}   &  \small{\multirow{2}{*}{$31.2 \times 10^{-8}$}} \\
                        & \\

\hline
\small{\multirow{2}{*}{$\mathcal{B}(B \to \mybar{\Lambda}^{0} \mybar{\Omega}_{c}^{0})$}} &  \small{\multirow{2}{*}{$9.5 \times 10^{-8}$}} \\
                                 & \\

\hline
\small{\multirow{2}{*}{$\mathcal{B}(B \to \mybar{\Lambda}^{0} \mybar{\Omega}_{c}^{}(2770)^{0})$}}   &  \small{\multirow{2}{*}{$10.0 \times 10^{-8}$}} \\
                        & \\
\end{tabular}
\end{ruledtabular}
\caption{Summary of the results.
}
\label{table_results}
\end{table}
}

In summary, 
we use the full data sample recorded 
by the Belle experiment at the $\Upsilon(4S)$ resonance
to search for SM and BNV two-body decays of neutral $B$ mesons 
to $\Omega^{0}_c$ or $\Omega_c^{}(2770)^{0}$, and $\Lambda^{0}$.
We observe no statistically significant signals and set 95\% CL upper
limits in the range between $9.5 \times 10^{-8}$ and $31.2 \times 10^{-8}$
on the products of neutral $B$ meson branching fractions
to $\Lambda^0$ and $\Omega_c^{(*)0}$
with $\mathcal{B}(\Omega_{c}^{0} \to \pi^+ \Omega^-)$.
The analysis presented in this Letter is the first study of
such SM-compatible and exclusively-BSM decays.
This work, based on data collected using the Belle detector, which was
operated until June 2010, was supported by 
the Ministry of Education, Culture, Sports, Science, and
Technology~(MEXT) of Japan, the Japan Society for the 
Promotion of Science~(JSPS), and the Tau-Lepton Physics 
Research Center of Nagoya University; 
the Australian Research Council including grants
DP210101900, 
DP210102831, 
DE220100462, 
LE210100098, 
LE230100085; 
Austrian Federal Ministry of Education, Science and Research~(FWF) and
FWF Austrian Science Fund No.~P~31361-N36;
National Key R\&D Program of China under Contract No.~2022YFA1601903,
National Natural Science Foundation of China and research grants
No.~11575017,
No.~11761141009, 
No.~11705209, 
No.~11975076, 
No.~12135005, 
No.~12150004, 
No.~12161141008, 
and
No.~12175041, 
and Shandong Provincial Natural Science Foundation Project ZR2022JQ02;
the Czech Science Foundation Grant No. 22-18469S;
Horizon 2020 ERC Advanced Grant No.~884719 and ERC Starting Grant No.~947006 ``InterLeptons''~(European Union);
the Carl Zeiss Foundation, the Deutsche Forschungsgemeinschaft, the
Excellence Cluster Universe, and the VolkswagenStiftung;
the Department of Atomic Energy~(Project Identification No. RTI 4002), the Department of Science and Technology of India,
and the UPES~(India) SEED finding programs Nos. UPES/R\&D-SEED-INFRA/17052023/01 and UPES/R\&D-SOE/20062022/06; 
the Istituto Nazionale di Fisica Nucleare of Italy; 
National Research Foundation~(NRF) of Korea Grant
Nos.~2016R1\-D1A1B\-02012900, 2018R1\-A2B\-3003643,
2018R1\-A6A1A\-06024970, RS\-2022\-00197659,
2019R1\-I1A3A\-01058933, 2021R1\-A6A1A\-03043957,
2021R1\-F1A\-1060423, 2021R1\-F1A\-1064008, 2022R1\-A2C\-1003993;
Radiation Science Research Institute, Foreign Large-size Research Facility Application Supporting project, the Global Science Experimental Data Hub Center of the Korea Institute of Science and Technology Information and KREONET/GLORIAD;
the Polish Ministry of Science and Higher Education and 
the National Science Center;
the Ministry of Science and Higher Education of the Russian Federation
and the HSE University Basic Research Program, Moscow; 
University of Tabuk research grants
S-1440-0321, S-0256-1438, and S-0280-1439~(Saudi Arabia);
the Slovenian Research Agency Grant Nos. J1-9124 and P1-0135;
Ikerbasque, Basque Foundation for Science, and the State Agency for Research
of the Spanish Ministry of Science and Innovation through Grant No. PID2022-136510NB-C33~(Spain);
the Swiss National Science Foundation; 
the Ministry of Education and the National Science and Technology Council of Taiwan;
and the United States Department of Energy and the National Science Foundation.
These acknowledgements are not to be interpreted as an endorsement of any
statement made by any of our institutes, funding agencies, governments, or
their representatives.
We thank the KEKB group for the excellent operation of the
accelerator; the KEK cryogenics group for the efficient
operation of the solenoid; and the KEK computer group and the Pacific Northwest National
Laboratory~(PNNL) Environmental Molecular Sciences Laboratory~(EMSL)
computing group for strong computing support; and the National
Institute of Informatics, and Science Information NETwork 6~(SINET6) for
valuable network support.


\begin{thebibliography}{99}
\bibitem{cc}
{The inclusion of charge-conjugate decay modes is implied throughout this Letter, unless stated otherwise.}
\end{thebibliography}
\end{document}